\documentclass[iop]{emulateapj}

\def\lapp{\ifmmode\stackrel{<}{_{\sim}}\else$\stackrel{<}{_{\sim}}$\fi}
\def\gapp{\ifmmode\stackrel{>}{_{\sim}}\else$\stackrel{>}{_{\sim}}$\fi}
\usepackage{multirow}
\usepackage{color}
\usepackage{amsmath}
\usepackage{soul}
\usepackage{hyperref}
\usepackage{amsmath}

\begin{document}

\title{SED constraints on the highest-$z$ blazar jet: QSO~J0906$+$6930}

\author{
Hongjun An\altaffilmark{1,*}, and Roger W. Romani\altaffilmark{2}
\\
{\small $^1$Department of Astronomy and Space Science, Chungbuk National University, Cheongju, 28644, Republic of Korea}\\
{\small $^2$Department of Physics/KIPAC, Stanford University, Stanford, CA 94305-4060, USA}\\
}
\altaffiliation{$^*$hjan@chungbuk.ac.kr}

\begin{abstract}
	We report on Gemini, {\it NuSTAR} and 8-year {\it Fermi}
observations of the most distant blazar QSO~J0906$+$6930 ($z=5.48$).
We construct a broadband spectral energy distribution (SED) and
model the SED using a synchro-Compton model.
The measurements find a $\sim 4 \times 10^9 M_\odot$ mass for the black hole 
and a spectral break at $\sim$4\,keV in the combined fit of the new {\it NuSTAR}
and archival {\it Chandra} data. The SED fitting constrains the bulk Doppler factor
$\delta$ of the jet to $9^{+2.5}_{-3}$ for QSO~J0906$+$6930.
Similar, but weaker $\delta$ constraints are derived from SED modeling of 
the three other claimed $z>5$ blazars. Together, these extrapolate to 
$\sim620$ similar sources, fully 20\% of the optically bright, high
mass AGN expected at $5<z<5.5$. This has interesting implications for
the early growth of massive black holes.
\end{abstract}


\section{Introduction}

	The existence of massive black holes (BH) at $z>5$ is well known, via
optical/IR surveys for bright quasars \citep[e.g.][]{fnls+01,mwvp+11}.
The most massive high-redshift sources present a puzzle;
it is challenging to grow a stellar mass seed black hole to 
$>10^9M_\odot$ levels in the limited age of the universe.  By inferring 
the cosmic density of quasars we can probe viable growth scenarios \citep[][]{bv08}. 

	Super-massive BHs can grow by merging or accretion.
Merging BHs may have random spin orientation and thus modest final BH
angular momentum.  Disk accretion growth increases the angular momentum 
along with the mass, but the accretion energy yield also increases
with the spin, so that Eddington-limited mass growth rates would decrease.
Thus large accretion-fed masses may be particularly difficult
to reach at early times. The so-called blazars are active galaxies dominated by the
two-humped (synchrotron + Compton) spectral energy distribution (SED) of relativistic jet emission. As such
they are bright microwave-IR (synchrotron) and gamma-ray (Compton) sources.
Moreover it is believed that this large jet power can be traced to
efficient extraction of rotational energy of a black hole with spin $a$ \citep{bz77}.
Thus searches for blazar sources at high $z$ \citep[see also][]{aabb+17} may
be a particularly interesting probe of the accretion-dominated
growth channel.

        Inspired by early {\it EGRET} detections, radio/optical surveys have indeed
found many blazars \citep{srmh+05,hrts+07}; most have now been detected by 
{\it Fermi} and have led to better understanding of the evolution of this massive,
jet-dominated BH population \citep{arg+14}. But these objects are largely
at relatively modest $z<3$ redshifts.  To date only four blazars
at $z>5$ have been reported in the literature
\citep[i.e., QSO~J0906$+$6930, B2~1023+25, SDSS~J114657.79+403708.6, 
SDSS~J0131$-$0321;][]{rsgp04,stgp+13,gstf+14,gtsg15}. Although we do not have formal
evaluation of the completeness of this sample, the sources are identified from
wide area radio surveys (\S 3), and therefore, the study of these objects can help us
understand the high redshift blazar population.

	This small sample size may be natural since the emission is dominated
by the relativistic jet, which is highly beamed. For bulk Lorentz factor 
$\Gamma_{\rm D}$, each blazar detection represents $\sim 2\Gamma_{\rm D}^2$ 
similar sources beamed away from Earth (for a more detailed estimate see \S 4).
Thus population inferences require careful extrapolation
of these few detected sources with good estimates of the viewing angle ($\theta_{\rm V}$) 
and bulk Doppler factor $\delta=1/[\Gamma_{\rm D}(1-\beta\mathrm{cos} \theta_{\rm V})]$, where
$\beta=\sqrt{1-1/\Gamma_{\rm D}^2}$. 
These can be extracted by measuring the SED
using the different dependencies of the synchrotron ($\nu_{sy, pk}^{obs}\propto \delta$),
self-Compton ($\nu_{ssc, pk}^{obs}\propto \delta$, SSC)
and external Compton ($\nu_{pk}^{obs}\propto \delta^2$, EC) components.
Adequately defining these emission components is, however, particularly challenging, since 
the known  $z>5$ blazars lack {\it Fermi} detections, and their synchrotron emission 
peak seems to fall in the millimeter wavelength range. We therefore rely on
hard X-ray measurements and GeV upper limits to constrain $\delta$.

	The most distant ($z=5.48$) blazar is the radio bright $S_{8.4 GHz} \sim 140$\,mJy 
GB6 0906+6930 (hereafter Q0906). It was actually found 
coincident with a low-significance ($1.5\sigma$ at one epoch) excess of {\it EGRET} 
gamma rays. \citet{rsgp04} and \citet{r06}
measured the SED of Q0906 in the radio to X-ray band and generated models that
could allow the {\it EGRET} detection. These implied $\Gamma_{\rm D}\sim 13$,
but the SED peaks were not well constrained. New {\it Fermi} upper
limits presented here imply substantially lower average gamma-ray flux.
Given that large $\Gamma_{\rm D}$, and similar values inferred for other
$z>5$ blazars would suggest large accretion-fed populations, improved
constraints on the blazar properties are needed.

	Here, we report on Gemini, {\it NuSTAR} and 8-yr {\it Fermi}-LAT observations and
jet properties inferred by SED modeling.
We describe the broadband data we collect in Section~\ref{sec:sec2} and report the 
data analysis results and SED modeling in Section~\ref{sec:sec3}. We then discuss
implications of our studies and conclude in Section~\ref{sec:sec4}.
We use $H_0=70\rm \ km\ s^{-1}, \Omega_m=0.3$ and $\Omega_\Lambda=0.7$
throughout \citep[][]{ksd+11}.

\section{Observation Data and Basic Processing}
\label{sec:sec2}

	In the IR band, Q0906 was observed with the Gemini-N GNIRS on December 3, 2015
(program GN-2015B-FT-22), using the short $0.15^{\prime\prime}$/pixel camera, the 32l/mm 
grating and the $0.3^{\prime\prime}$ slit at the average parallactic angle. This 
provides coverage from 0.88--2.5$\mu m$ in orders 3 through 8 with resolving power
$R\sim 1400$. Relative calibration was provided by $4\times 2$\,s spectroscopic 
integrations (ABBA pattern) of the flux standard HIP43266. Although the standard was
acquired with a direct $H$ image, this was saturated, so we could not measure the
slit losses to establish an absolute flux scale. Next Q0906 was observed, starting with
three direct images through the $H$ filter, each comprised of $12\times 3$\,s co-added.
The stacked image FWHM was $0.41^{\prime\prime}$. We then obtained 12 spectroscopic 
exposures of 300\,s, dithering along the slit in an ABBA pattern. The first two
exposures suffered contamination from a bright persistence signal, while the last
two had a severely increased background from morning twilight. This left 8 useful
exposures, totaling 2400\,s.

	We also observed Q0906 with the {\it NuSTAR} observatory \citep{hcc+13}
between MJD~57732 (2016-12-10 UTC) and 57734 (2016-12-11 UTC) with 75\,ks 
total exposure (LIVETIME) to collect a hard X-ray spectrum in the 3--79\,keV band.
For this observation, the {\it NuSTAR} Science Operation Center (SOC) reported
slightly elevated background rates
around South Atlantic Anomaly (SAA) passage and recommended that we use more strict
filters. The data are downloaded from the {\it NuSTAR} archive and are
processed with the {\tt nupipeline} tool integrated in
{\tt HEASOFT}~6.19 along with the {\it NuSTAR} CALDB (release 20160706) with
the strict filters suggested by the SOC.

        In the gamma-ray band, we use the Pass-8 reprocessed {\it Fermi}-LAT data \citep{fermimission,fermiP8}
collected between 2008 August 04 and 2016 November 28 UTC.
We processed the data with the {\it Fermi}-LAT Science Tools {\tt v10r0p5}
along with P8R2\_V6 instrument response functions,
and selected source class events with Front/Back event type in an
$R=10^\circ$ aperture in the 100\,MeV--300\,GeV band. We further employed standard
$<90^\circ$ zenith angle and $52^\circ$ rocking angle cuts.

	Broadband coverage helps us pin down the SED peaks and so we use
archival radio, IR, optical and
soft X-ray data. For the IR data, we take the measurements from the {\it WISE} and the {\it Spitzer}
catalogs. For the radio and the optical data, we use
the measurements reported by \citet{r06}. In the soft X-ray band ($<$10\,keV), we reanalyze the
archival 30-ks {\it Chandra} data \citep{r06} and the {\it Swift}/XRT data (11 exposures).
The {\it Chandra} data are reprocessed with {\tt chandra\_repro} of CIAO~4.8 using
CALDB~4.7.2, and the {\it Swift} data are processed with {\tt xrtpipeline}
in HEASOFT~6.19 with the HEASARC remote CALDB.
Note that the latest {\it Swift} exposure is contemporaneous with the {\it NuSTAR}
observation; comparison with other {\it Swift} epochs confirm that the blazar was in an average
state, suitable for comparison with non-simultaneous multiwavelength observations.

\section{Data Analysis and Modeling}
\label{sec:sec3}

\subsection{Gemini Data Analysis}
\label{sec:sec3_3}
\begin{figure}
\centering
\hspace{-4 mm}
\includegraphics[width=3.5 in]{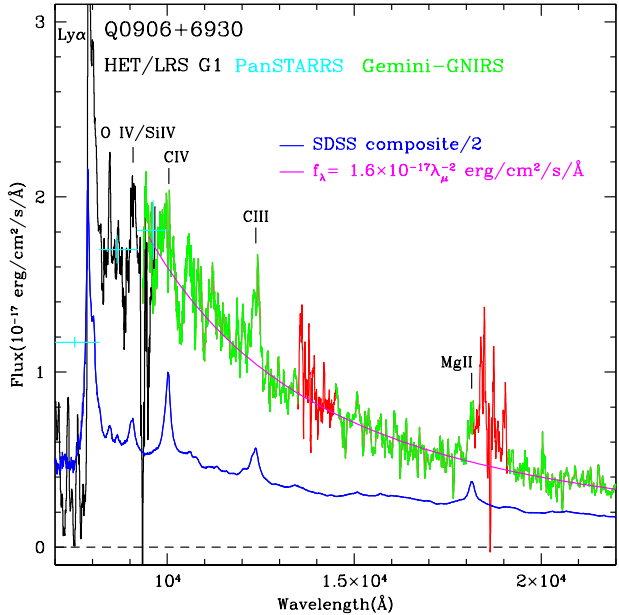} 
\figcaption{The near-IR spectrum of Q0906, with the HET G1 (black) and
GNIRS (green) spectra. These are matched to imaging PanSTARRS fluxes
($i$, $z$ and $y$ shown as red error flags).
In the GNIRS spectrum, regions between the near-IR windows have
very low S/N and are plotted in red.  A redshifted composite SDSS QSO spectrum is shown
in blue, scaled down by $2\times$ for visibility. A simple power-law approximation to the GNIRS
continuum is plotted in magenta.  Strong UV resonance lines are marked. C IV
is weak and appears affected by poor continuum fluxing. Mg II is partly lost
to atmospheric absorption.
\label{fig:fig1}
}
\vspace{0mm}
\end{figure}

	The GNIRS spectra were reduced with scripts from the Gemini 1.13 package, including
removal of pattern noise, flat fielding, rectification, wavelength calibration
and fluxing with the standard spectrum. The orders were combined into a single spectrum,
which is smoothed and plotted in Figure~\ref{fig:fig1}. The unsmoothed S/N/pixel peaks at $\sim 6$ in 
the middle of the orders. The J/H (1.35--1.45$\mu m$) and H/K (1.82--1.91$\mu m$)
gaps, with low atmospheric transmission, are particularly noisy and are plotted in red.
We also show the HET G1 spectrum of \citet{rsgp04}. With unknown differential slit
losses between the standard and target exposures, we elect to normalize to the direct
imaging fluxes measured by PanSTARRS\footnote{http://archive.stsci.edu/panstarrs}. We integrate
the HET G1 spectrum over $0.818-0.922\mu m$ and scale to the magnitude $z=19.83\pm0.05$,
while an integration of the GNIRS spectrum over $0.918-1.001\mu m$ is scaled to the 
$y=19.54\pm0.09$ image flux.

        We examined our direct $H$ image where the quasar is well detected. The
GNIRS H-band spectrum has a flux of $5.9 \times 10^{-18} {\rm erg/cm^2/s/\AA }$. No other
source is detected in the GNIRS `keyhole' field, placing an upper limit
of $3.5\times 10^{-19} {\rm erg/cm^2/s/\AA }$ on any source within $\sim 3^{\prime\prime}$
of the quasar (limited by the nearby edge of the keyhole field of view).
This provides a modest limit of $L<2.3 L_\ast$ on the luminosity of any intervening
galaxy associated with the strong Mg II absorption system at $z=1.849$ \citep{r06}.
No nearby sources are seen in the PanSTARRS images.

        The IR spectrum has a continuum approximated by
$f_\lambda = 1.6 \times 10^{-17} \lambda_\mu^{-2} {\rm erg/cm^2/s/\AA}$.
In Figure~\ref{fig:fig1} we also plot the SDSS composite QSO spectrum of \citet{vrbs+01},
redshifted to $z=5.48$ and scaled down by $\sim 2 \times$.
We are particularly interested in the UV emission lines shifted to the IR.
C IV (1550) is rather poorly detected, being absorbed by
a strong (rest frame EW=0.8\AA) associated doublet at $z=5.469$ and being flanked by large continuum
oscillations, possibly due to poor fluxing.  Its overall weakness is likely a consequence
of the large QSO luminosity (Baldwin effect).  C III (1909) is well detected, while
Mg II (2800) is at the edge of the H-band with the red half of the line lost to
atmospheric absorption. This is unfortunate, since we can use neither of the standard
calibrated species (C IV, Mg II) for a virial mass estimate. For the C III line
we measure a Gaussian FWHM=$6200 \pm 400$\,km/s. The left half of the Mg II line provides
a line width estimate FWHM$\approx 6000$\,km/s. For C IV the poorly defined continuum
prevents any meaningful line width estimate. In \citet{rsgp04} the O IV/S IV line
was estimated to have FWHM=$5000 \pm 500$\,km/s. If we assume a line width of 6000\,km/s
then, measuring the standard continuum luminosity for
C IV ($\lambda L_\lambda{1350} = 5.4 \times 10^{46} {\rm erg/s}$) gives log$M_\bullet=9.66$
\citep[][]{md04} while Mg II ($\lambda L_\lambda{3000} = 2.7 \times
10^{46} {\rm erg/s}$) gives log$M_\bullet=9.57$.
These give an average inferred BH mass $M_\bullet = 4.2 \times 10^9 M_\odot$, subject to the
usual systematic uncertainties as well as the errors from the
poor spectral line measurements. Still, these IR spectra provide a useful
mass estimate and show the flattening of the IR SED bump toward a peak at $\sim 0.9\mu m$.

\subsection{X-ray Data Analysis}
\label{sec:sec3_1}

	Because blazars are often variable in all wavebands,
combining data taken in different epochs needs to be done with care. We therefore first
checked for time variability of the X-ray flux using {\it Swift} archive
data, which have observations spanning 11 years (2006 Jan. 20--2016 Dec. 10).  
We constructed a long-term light curve using
circles with $R=20''$ and $R=40''$ for source and background extraction,
respectively. A total of $76\pm11$ background-subtracted events were
detected over the integrated 43\,ks exposure (summed over the 11-yr observations).
All epochs have count rates within 60\% of the average and there is no evidence
for spectral variability; within the statistic-limited sensitivity, the light 
curve is consistent with being constant. In particular, the {\it Swift} observation
contemporaneous with the {\it NuSTAR} observation, has a count rate consistent
with the light curve average and with the count rate expected from the measured
{\it Chandra} spectrum (see below). Therefore, we conclude that the Q0906 variability 
is small, that it was in a typical state during the {\it NuSTAR} exposure,
and that we can reasonably combine non-contemporaneous observations in
forming the SED.

	For X-ray spectral properties, we first reanalyzed the {\it Chandra} data. 
Source events were extracted from a $R=2''$ circular aperture and background from an
annular region with $R_{\rm in}=5''$ and $R_{\rm out}=10''$ centered at the
source position. Response files are calculated using the {\tt specextract} tool
of CIAO.  For the absorption model, we use $\tt wilm$ abundance \citep[][]{wam00} 
and $\tt verner$ cross section \citep[][]{vfky96}.
We group the spectrum to have at least 20 counts per bin and fit
the spectrum with an absorbed power-law model (PL) in XSPEC. The spectral parameters are
consistent with previous measurements \citep[][]{r06}. 

	The 11 {\it Swift} observations were separately analyzed.
We used a $R=20''$ circular region for source spectra. A source-free nearby
$R=40''$ region was used for background extraction. The ancillary response files (ARFs)
were produced with the {\tt xrtmkarf} tool correcting for the exposure, and we
used pre-generated redistribution matrix files (RMFs). After this, we find that
individual {\it Swift} spectra do not have enough events for a meaningful spectral
analysis. We therefore combine all
the spectra with the {\tt addspec} tool of HEASOFT. We grouped the
combined spectrum to have at least 20 counts per spectral bin, and fit the spectrum
with an absorbed power-law model holding $N_{\rm H}$ fixed at the {\it Chandra}-measured
value.
Employing different statistics (e.g., {\tt lstat} in XSPEC) or different binning
does not change the parameters significantly.
The results are also consistent with the {\it Chandra} values above (Table~\ref{ta:ta1}).

\begin{figure*}
\centering
\hspace{-7 mm}
\begin{tabular}{cc}
\includegraphics[width=3.5 in]{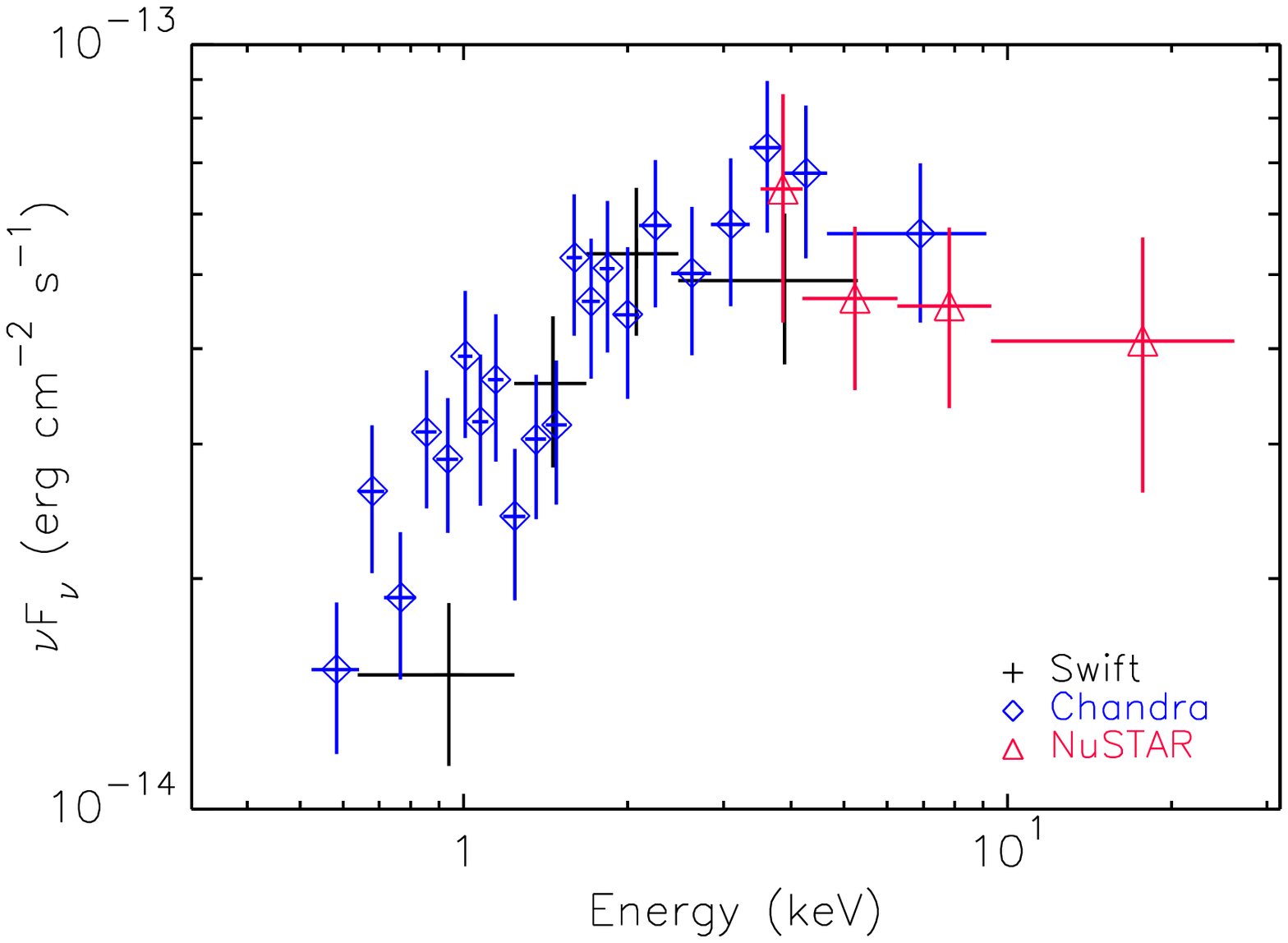} &
\includegraphics[width=3.5 in]{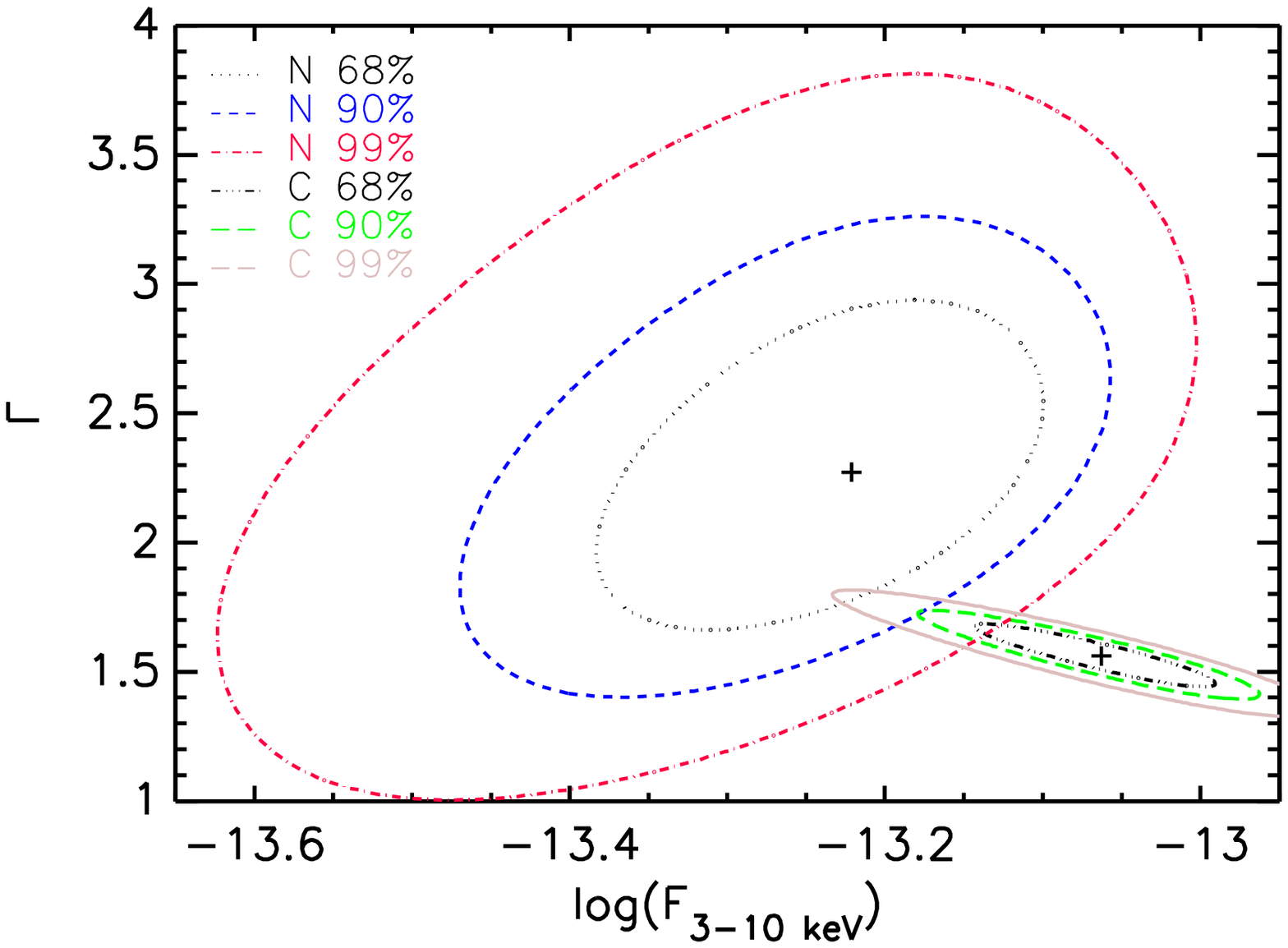} \\
\end{tabular}
\figcaption{{\it Left}: X-ray SED of Q0906 measured with {\it Swift}, {\it Chandra} and {\it NuSTAR}.
{\it Right}: confidence contours in the $\mathrm logF$-$\Gamma$ space for the {\it NuSTAR}
and {\it Chandra} fits.
68\%, 90\%, and 99\% contours corresponding to $\Delta \chi^2=$2.3, 4.61 and 9.21, respectively,
are shown in lines.
\label{fig:fig2}
}
\vspace{0mm}
\end{figure*}

	For the {\it NuSTAR} data analysis, we used $R=30''$ and 
$R=60''$ regions for the source and the background extraction, respectively.
The corresponding response files were calculated with the {\tt nuproduct} tool.
Q0906 was surprisingly faint, yielding only 120$\pm$20 source
events in the 3--20\,keV band while we expected 260 counts if the {\it Chandra}-measured power-law spectrum
extends to higher energies. Keeping this in mind, we grouped the spectra to have
at least 20 events per spectral bin and fit the {\it NuSTAR} spectra with
a simple power-law model. We find that the measured 3--20\,keV spectrum is softer
than the soft band determination (Table~\ref{ta:ta1}), having an index
$\Gamma=2.3\pm0.4$ (Figure~\ref{fig:fig2} left). For such a faint source, the fit parameters might be sensitive
to the fit statistic or background selection. We therefore varied both;
we used three different $R=60''$ background regions,
and fit the data using $l$ statistic or $\chi^2$ statistic.
None of these tests gave significant changes to the measured parameters.

	The low {\it NuSTAR} count rate may imply a softer spectrum at higher energies.
So we checked to see if the best-fit {\it NuSTAR} parameters are consistent
with the {\it Chandra}-measured values using the {\tt steppar} command of XSPEC.
This provides confidence contours for the {\it NuSTAR} parameters and
shows that the {\it Chandra} values lie outside the 99\% contour (Figure~\ref{fig:fig2} right),
although including the {\it Chandra} parameter uncertainty, there is some overlap.

\newcommand{\markaa}{\tablenotemark{a}}
\newcommand{\markbb}{\tablenotemark{b}}
\begin{table*}[t]
\vspace{-0.0in}
\begin{center}
\caption{X-ray spectral fit results for Q0906}
\label{ta:ta1}
\vspace{-0.05in}
\scriptsize{
\begin{tabular}{ccccccc} \hline\hline
\multicolumn{7}{c}{Individual fits} \\
Instrument	& Model	& $\Gamma_{\rm s}$\markaa	& $E_{\rm b}$\markaa &$\Gamma_{\rm h}$\markaa	&$F$\markbb  & $\chi^2$/dof \\ 
		& 	&		& (keV) 		&	   &  \\ \hline
S  	& PL	& $1.32\pm0.21$	& $\cdots$	& $\cdots$ 	&	$1.5\pm0.3$  &  4/4	 \\
C  		& PL	& $1.56\pm0.14$	& $\cdots$	& $\cdots$ 	&	$1.5\pm0.1$  &  14/18	 \\
N  	& PL	& $2.27\pm0.41$	& $\cdots$	& $\cdots$ 	&	$0.6\pm0.1$  &  8/11	 \\ \hline
\multicolumn{7}{c}{Joint fits} \\ 
S+C+N  	& PL	& $1.55\pm0.07$	& $\cdots$	& $\cdots$ 	&	$0.46\pm0.10$  &  30/34 \\ 
S+C+N  	& BPL	& $1.44\pm0.10$	& $3.8\pm1.0$	& $2.33\pm0.42$ 	&	$0.60\pm0.13$ & 25/32 \\ \hline
\end{tabular}}
\end{center}
\vspace{-0.5 mm}
\footnotesize{
{\bf Notes.} $N_{\rm H}$ is measured to be $8\times 10^{20}\rm \ cm^{-2}$ with the
{\it Chandra} fit and held fixed at this value for the other fits. Instruments are
{\it Swift} (S), {\it Chandra} (C), and {\it NuSTAR} (N).\\}
$^{\rm a}${Power law index $\Gamma_{\rm s}$. If broken at  $E_{\rm b}$ hard index is $\Gamma_{\rm h}$.}\\
$^{\rm b}${Absorption-corrected 0.5\,keV--10\,keV flux in units of $10^{-13}\rm \ erg\ cm^{-2}\ s^{-1} $
for {\it Swift} and {\it Chandra} fits and 3\,keV--10\,keV flux for {\it NuSTAR} and joint fits.}\\
\end{table*}

	A joint XSPEC fit of the {\it Chandra}, {\it Swift}
and {\it NuSTAR} data to a simple power law with free cross-normalization
yields $\Gamma=1.55\pm0.07$ consistent with that measured with {\it Chandra} 
alone. This is evidently due to the {\it Chandra} count dominance. The fit tension is
revealed in the anomalously large cross-normalization factor $1.9\pm0.6$ with {\it Chandra}.
If we fit the data sets
with an absorbed broken power law (BPL), the cross-normalization factor for {\it Chandra} becomes 
$1.2\pm0.3$, consistent with the nominal calibration offset of $\sim$10\% \citep[][]{mhma+15}.
The best-fit parameters for this broken power-law model are $\Gamma_{\rm s}=1.44\pm0.10$,
$E_{\rm break}=3.8\pm1.0$\,keV, and $\Gamma_{\rm h}=2.3\pm0.4$ (see Table~\ref{ta:ta1}).
The improvement of these broken power-law fits is modest, but the improved relative
normalization lends confidence that this is a better model.
Note that similar spectral breaks have been seen in other blazars \citep[][]{hnms+15,tgph+15,sgtp+16,ppfs16}
and were variable in some cases.

	Although we do not significantly detect the source above $\sim$30\,keV, {\it NuSTAR}
can still be used to derive an upper limit that gives a useful constraint on the
SED Compton peak. Fixing the index at the $\Gamma_{\rm h}=2.33$ of the broken power-law
fit, we use the {\tt steppar} tool of XSPEC to scan the normalization while comparing
with the 20--79\,keV {\it NuSTAR} data. Finding the value at which  $\Delta \chi^2$ 
increases by 2.71, we establish a 95\% flux upper limit of
$2.4\times 10^{-13}\rm \ erg\ cm^{-2}\ s^{-1}$.
The SED is shown in Figure~\ref{fig:fig3} (top left).

\subsection{Fermi-LAT Data Analysis}

\label{sec:sec3_2}

We next derived a spectrum from 8-yr years of
100\,MeV--300\,GeV {\it Fermi}-LAT `Pass8' data using binned likelihood
analysis\footnote{https://fermi.gsfc.nasa.gov/ssc/data/analysis/documentation\\/Pass8\_usage.html}.
We fit the spectrum to a power-law model using the {\tt pyLikelihood} package provided along
with the Science Tools. Because Q0906 is not in the 3FGL catalog \citep[][]{fermi3fgl}, we
added it to the 3FGL {\tt XML} model assuming a power-law spectrum.
We then fit the data, varying parameters for Q0906, nearby bright sources,
the diffuse emission \citep[{\tt gll\_iem\_v06.fits};][]{fermigllv06} and the isotropic emission
\citep[{\tt iso\_P8R2\_SOURCE\_V6\_v06.txt};][]{fermiiso} in the 100\,MeV--300\,GeV band.
Q0906 is not detected, having a mission-averaged test statistic (TS) value of $<1$.
We also varied the number of nearby sources to fit
and the aperture size (Region of interest, RoI $R=5^\circ$ and $R=15^\circ$), and
found that the result does not change. We therefore report the 95\% flux upper
limit for Q0906 of $6\times 10^{-10}\rm \ ph\ cm^{-2}\ s^{-1}$ in the 100\,MeV--300\,GeV band
assuming a typical $\Gamma=2$ photon index derived using the {\tt UpperLimits.py} script,
which scans the power-law amplitude to find the value for which the loglikelihood
($-\mathrm log\mathcal L$) increases by 1.35 from the minimum value.

\begin{figure*}
\centering
\hspace{-0 mm}
\includegraphics[width=7.0 in]{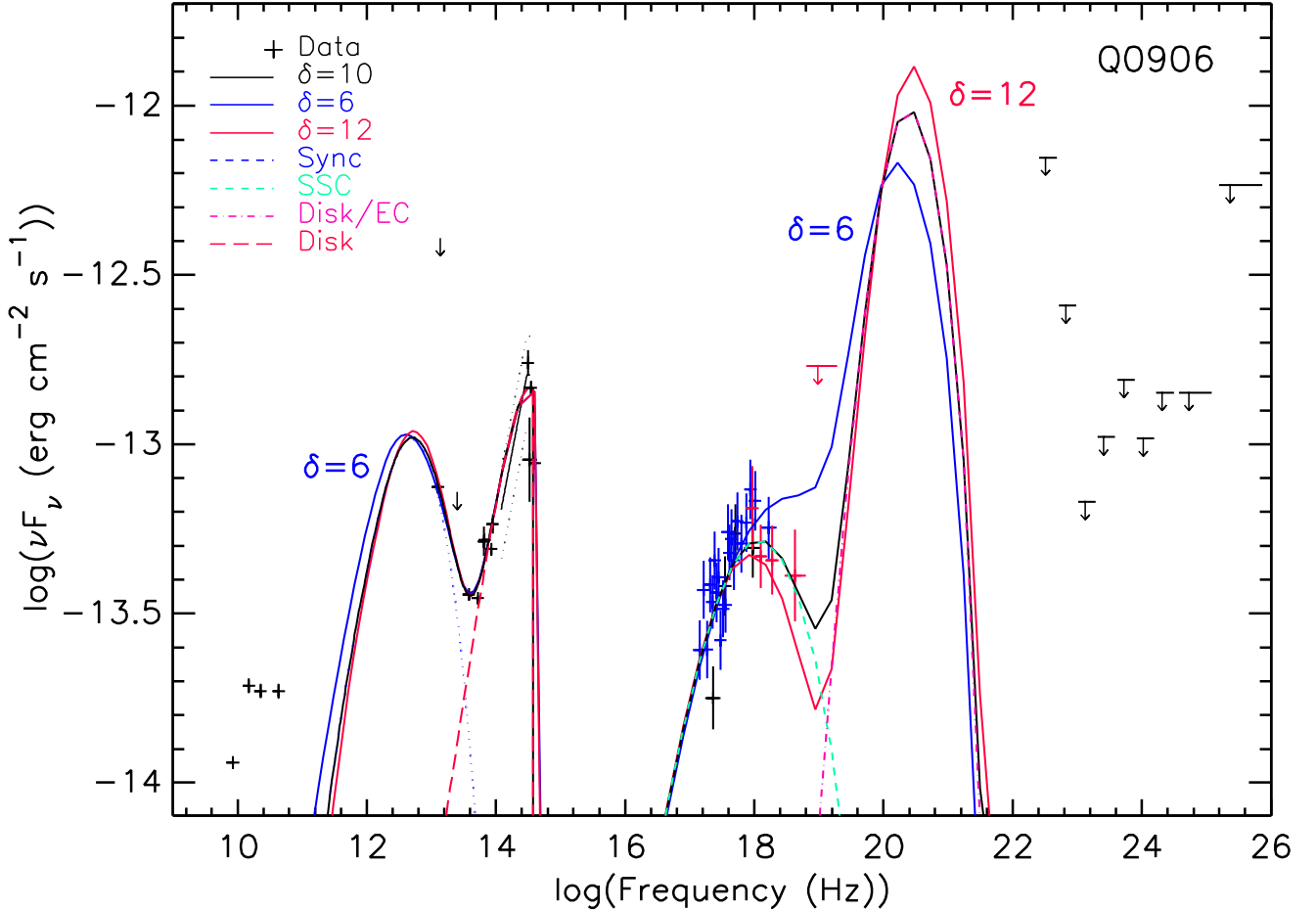}
\figcaption{The broadband SED of Q0906 and the synchro-Compton models.
The model parameters are adjusted to match the SEDs and shown in Table~\ref{ta:ta2}.
We also show two models with a lower (blue) and the maximum (red) $\delta$:
6 and 12 for Q0906. We show the X-ray data in color for clarity:
black for {\it Swift}, blue for {\it Chandra} and red for {\it NuSTAR}.
\label{fig:fig3}
}
\end{figure*}

	Next, since the strongest SED constraints may be energy dependent,
we derive the LAT SED of Q0906 in nine energy bands.
For this, we assumed a power-law spectrum across each band with $\Gamma=2$ and fit the amplitude
of the power-law model in individual energy bands.
As expected, the source was not detected (TS$<$9) in any of the energy bands, and we provide the
95\% flux upper limits. These gamma-ray flux limits shown in Figure~\ref{fig:fig3} (top left).
We note that the upper limits are not very sensitive to the assumed power-law index.

\begin{figure*}
\centering
\hspace{0 mm}
\includegraphics[width=7. in]{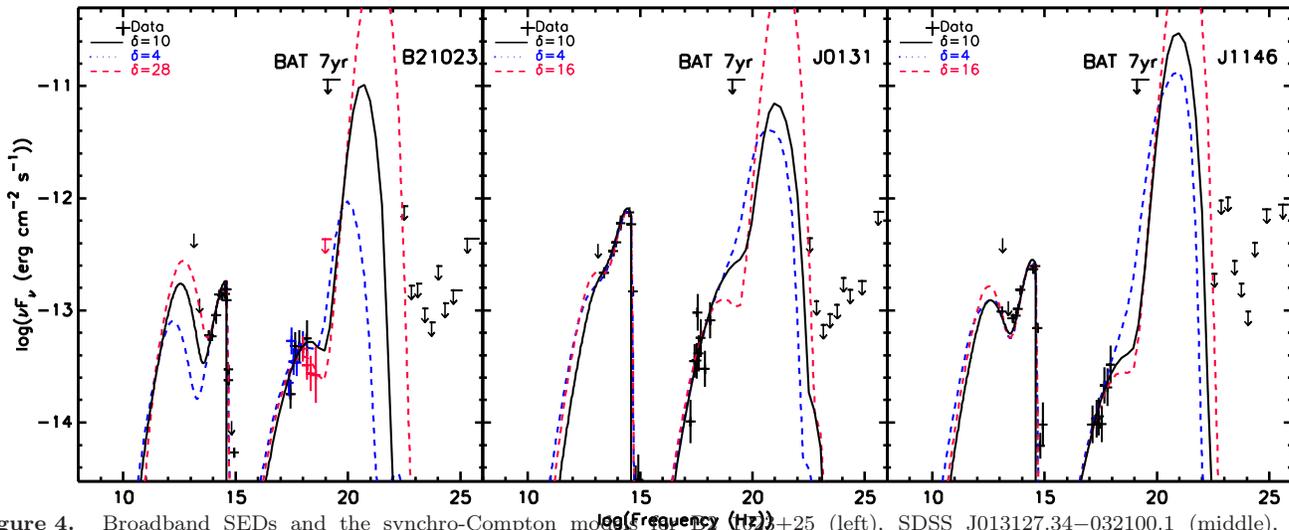}
\vspace{-5 mm}
\figcaption{Broadband SEDs and the synchro-Compton models for
B2~1023+25 (left), SDSS~J013127.34$-$032100.1 (middle),
and SDSS~J114657.79+403708.6 (right).
Models with a lower (blue dotted) and the maximum (red dashed) $\delta$,
4 and 28 for B2~1023, 4 and 16 for J0131, and 4 and 16 for J1146 are also shown.
For B2~1023+25, the same models for Q0906 work well but here we show a different set
of models with lower-$B$ and larger EC emission; this is possible for B2~1023 because
the synchrotron SED is not constrained with the data.
\label{fig:fig4}
}
\end{figure*}

	Finally, we check to see if the source is variable in gamma rays. In particular,
if there had been a large flare, the source might have had higher significance
during a restricted period. For this, we generated
a light curve using 1-Ms time bins and performed likelihood analysis to derive 100\,MeV--300\,GeV
flux in each time bin.
In the likelihood analysis, we vary the amplitudes of Q0906 and
nearby bright and variable sources (with the variability index greater than 100 in the 3FGL catalog).
The test statistics for Q0906 is less than 9 in most of the time intervals.
There is one time interval in which the detection significance is higher
(TS$\approx$12, MJD~56396--56407).
Although this is the highest value we get in our analysis, the probability of having
such a value or greater in 263 trials (time bins) is 14\%,
implying that this is not sufficient to claim a detection.

\subsection{Broadband SED Modeling}
\label{sec:sec3_4}

	Combining the new and archival data (Sections~\ref{sec:sec3_1} and \ref{sec:sec3_2})
we construct the broadband SED for Q0906 in Figure~\ref{fig:fig3}.
From the SED peak frequencies, we can derive rough constraints on the bulk Doppler factor. 
In synchro-Compton models, the low energy hump in the SED is produced by synchrotron radiation.
Thermal emission from the disk provides an intermediate peak 
at $\nu_{BB,pk}$ in the IR-optical band \citep{ss73} for Q0906.
X-ray emission is produced by self-Compton scattering of the synchrotron emission, 
and an external Compton component from up-scattered disk photons will produce a
hump in the MeV band (see Figure~\ref{fig:fig3}). The SED peak frequencies are related by 
$\nu_{sy,pk}^{obs} \approx \frac{\nu_{sy,pk} \delta}{1+z}$ (Synchrotron, with
\noindent$\nu_{sy,pk} \approx 3.7\times10^6 \gamma_e^2 B$), 
$\nu_{ssc,pk}^{obs} \approx \frac{\nu_{sy,pk} \delta \gamma_e^2}{1+z} = \gamma_e^2 \nu_{sy,pk}^{obs}$
(self-Compton) and 
$\nu_{EC,pk}^{obs} \approx \frac{\nu_{BB,pk} \delta^2 \gamma_e^2}{1+z} = \delta^2 \gamma_e^2 \nu_{BB,pk}^{obs}$
(external inverse Compton).
With incomplete coverage, the peak frequencies are uncertain, but from
the SED shape we estimate $\nu_{sy,pk}^{obs}\approx4\times10^{12}$\,Hz, 
$\nu_{BB,pk}^{obs}\approx3\times10^{14}$\,Hz,
$\nu_{ssc,pk}^{obs}\approx10^{18}$\,Hz, and 
$2\times10^{19}\rm\ Hz < \nu_{EC,pk}^{obs} < 10^{22}\rm \ Hz$.
From visual estimates of the SED peak positions and the frequency scaling above we see that 
$\gamma_e \approx (\nu_{ssc,pk}^{obs}/\nu_{sy,pk}^{obs})^{1/2} \approx 500$
and $\delta \approx (\nu_{EC,pk}^{obs}/\nu_{BB,pk}^{obs})^{1/2}/\gamma_e \approx 0.6-13$.

	We can make better estimates by comparing the data with detailed SED models. We
use the synchro-self-Compton model developed by \citet{bms97} to describe the broadband blazar SED.
This model assumes a continuous injection into a $e^+/e^-$ blob at the jet base (at a height
$h$=0.03\,pc from the BH) and evolves the blob
over an interval of $10^7$\,s following radiative losses. The properties of the blob (injection spectrum,
bulk Doppler factor, magnetic field strength, and so forth) are prescribed, and the code delivers
the integrated emission spectrum.
As above, we assume that the low-energy $\lapp10^{13}$\,Hz emission is produced by synchrotron radiation.
The radio points are, as usual, well above the expectation of the core synchrotron component as
these represent the late time emission of the blobs after they flow to large (VLBI-scale)
radii \citep[e.g.,][]{cbka+10}.
The $\sim10^{14}$\,Hz SED is the disk blackbody emission, absorbed to the blue by the intragalactic
Lyman-$\alpha$ forest.  Two processes contribute to the X-ray emission:
synchro-self-Compton radiation and external Compton up-scattering of the disk photons.
As one moves to higher X-ray energies, the external Compton from the higher-frequency
disk photons should become increasingly important. Since larger $\delta$ shifts the EC peak
to higher frequency, its contribution to the X-ray band is very sensitive to this factor.
For small $\delta$ we expect the sharp rise to the EC peak to enter the {\it NuSTAR} band;
for larger $\delta$ we will see the falling spectrum above the isolated $\nu_{ssc,pk}^{obs}$ peak.

	We compute the disk blackbody emission with a Shakura-Sunyaev \citep[][]{ss73} model.
In Figure 1, we appear to detect a continuum flattening above $\sim 3 \times 10^{14}$\,Hz.
However, the onset of Lyman-$\alpha$ forest absorption at $3.8 \times 10^{14}$\,Hz precludes
detailed measurement of the thermal peak. We therefore used the viral BH mass estimate
$M_\bullet=4.2 \times 10^{9}M_\odot$ (Section~\ref{sec:sec3_1}) and adjust $L_{\rm disk}$
to match the disk IR flux \citep[e.g.,][]{cgcd13}. The optical-IR SED matches that of a 
Shakura-Sunyaev disk for disk luminosity $L_{\rm disk}=2.4 \times 10^{47}\rm \ erg\ s^{-1}$.
This is $\sim 0.4 L_{Edd}$, suggesting that the thin disk approximation is adequate.
The virial estimates are uncertain and smaller BH masses adjust the disk luminosity:
e.g., $M_\bullet=3\times10^{9}M_\odot$ implies $L_{\rm disk}=2.6\times 10^{47}\rm \ erg\ s^{-1}$.
However, this uncertainty induces rather small ranges in the other model parameters,
so we neglect it below.

	 For blazars, the synchrotron-producing electron spectrum typically has an index
$p_1\approx2$; here we use $p_1=1.8$ for the best SED match. This spectrum ranges
from minimum $\gamma_{\rm e,min}$ to maximum $\gamma_{\rm e,max}$. These values and
the magnetic field strength $B$ are adjusted to match the shapes and the amplitudes of the synchrotron
and the X-ray SEDs. Given the large number of parameters, SED data alone are insufficient
to force unique values for each quantity. By assuming magnetic field equipartition,
we find $\gamma_{\rm e}\sim 5\times 10^{2}$ from 
$\nu_{\rm sy,obs} = 3.7\times10^6 \gamma_e^2 B\delta/(1+z)\sim 4\times 10^{12}\rm \ Hz$.
Note that electron power injected into the jet represents 
$\sim 10$\% of the thermal (disk) flux; beaming is what makes the jet dominate along the 
Earth line-of-sight. As described above this Doppler beaming also shifts the SED peaks. The EC peak
in particular is sensitive to $\delta$, with the  20--79\,keV X-ray and LAT upper limits setting
the allowed range. In Table~\ref{ta:ta2} we give the model parameters 
for $\delta=10$ in the middle of this range, and Figure~\ref{fig:fig3} shows example models
with values at the high and low extremes.

	In summary, large values of $\delta$ tend to push the external Compton peak to 
higher frequency. If too large this would violate the {\it Fermi}-LAT upper limits.
For small $\delta$, the low frequency side of the external Compton peak can
over-predict the {\it NuSTAR} measurement.  In practice the more detailed SED modeling
gives stronger constraints from the relative positions and fluxes of the peaks, including
the SSC peak in the X-rays. For example
our upper bound on $\delta$ arises from comparison of the synchrotron and SSC component amplitudes.
An increased $\delta$ boosts the SSC peak frequency and amplitude; to maintain
a data match for the SSC component we reduce both $B$ and $n_{\rm e}$ (maintaining equipartition).
But then the synchrotron flux $\propto n_{\rm e}$ drops slower than the SSC flux $\propto n_{\rm e}^2$,
and so is
over-produced. Reducing $\delta$ gives the opposite trend. With our new X-ray measurements
fixing the SSC peak, this constraint is particularly useful for Q0906.

	We search for the range of acceptable $\delta$ in the following way.
We first adjust the model parameters to match the SED for $\delta=10$. We further optimize
the model parameters using the Monte Carlo technique. We then change $\delta$ to a
different value (between 6 and 13), hold it fixed at the value, and adjust
the other parameters ($\gamma_{\rm e,min}$, $\gamma_{\rm e,max}$, $p_1$, $n_{\rm e}$, and $R'_b$)
to minimize $\chi^2$ for the synchrotron (the two lowest-frequency IR points)
and the SSC emission (X-ray points). The disk component is not considered in this minimization.
The fits match the IR points better by sacrificing the X-ray fits because of the small uncertainties
in the IR band. So the fits for different $\delta$'s differ mostly in the hard X-ray band.
We find that the X-ray $\chi^2$ has a minimum around $\delta=9$ ($\chi^2$/dof=36.8/25).
If we formally use the $\Delta \chi^2$ statistic for 6 parameters, we find $\delta=6-11.5$.
The extreme $\delta$ values are shown by the $\delta =6$ and $\delta =12$ lines
(Figure~\ref{fig:fig3}, upper left panel).

	We note that the $\sim 1.5\sigma$ EGRET `detection' described by \citet{rsgp04}
was for a single 2 week viewing period in 1992. The nominal flux would be substantially
higher than the {\it Fermi} upper limit in Figure~\ref{fig:fig3}. With {\it EGRET}'s
very soft response function, it is possible that this represents a brief low energy flare,
but it is more likely that this was just a statistical fluctuation and Q0906 remains
undetected in the gamma-ray band.

\newcommand{\marka}{\tablenotemark{a}}
\newcommand{\markb}{\tablenotemark{b}}
\newcommand{\markc}{\tablenotemark{c}}
\newcommand{\markd}{\tablenotemark{d}}
\begin{table*}[t]
\vspace{-0.0in}
\begin{center}
\caption{Parameters for the SED model}
\label{ta:ta2}
\vspace{-0.05in}
\scriptsize{
\begin{tabular}{cccccccc} \hline\hline
Parameter	& Symbol	& \multicolumn{4}{c}{Value}	\\ \hline
Target	 	&  	& Q0906	& B2~1023 & J0131 & J1146 \\ 
Redshift$^a$ 	& $z$	& 5.48	&  5.28 & 5.18 & 5.00 \\ 
Black Hole mass ($M_\odot^a$)& $M_\bullet$ & $4.2\times 10^9$ & $4\times 10^9$ & $1.5\times 10^{10}$ & $8\times 10^{9}$ \\ 
Disk Luminosity (erg/s)& $L_{disk}$	& $2.4\times 10^{47}$ & $2.7\times 10^{47}$ & $1.1\times 10^{48}$ & $3.5\times 10^{47}$ \\ 
Doppler factor 	& $\delta^b$	& 6--11.5 & 4--28 & 4--16 & 4--16 \\ 
Magnetic field (G)	& $B$	& 6.9 & 2.9 & 11 & 2.1 \\ \hline 
Comoving radius of blob (cm)	& $R'_b$& $8.8\times 10^{14}$ & $2.6 \times 10^{15}$ & $1\times 10^{15}$ & $4.1\times 10^{15}$ \\ 
Effective radius of blob (cm)\markc& $R'_E$ & $5.5\times 10^{15}$ & $1.1\times 10^{16}$ & $ 6\times 10^{15}$ & $ 1.6\times 10^{16}$ \\ 
Electron density (cm$^{-3}$)	& $n_e$ & $6.1\times 10^{3}$ & $7\times 10^{2}$ & $7\times 10^{3}$ &  $3\times10^{2}$ \\ 
Initial electron spectral index	& $p_1$ & 1.8 & 2.1 & 1.7 &  1.6 \\ 
Initial min. electron Lorentz factor	& $\gamma_{\rm min}$ & $2.2\times 10^{2}$ & $4\times 10^{2}$ & $4.5\times 10^{2}$ & $4.5\times 10^{2}$ \\ 
Initial max. electron Lorentz factor	& $\gamma_{\rm max}$ & $7.2\times 10^{2}$ & $10^{3}$ & $1.4\times 10^{3}$ & $1.2\times 10^{3}$ \\ 
Injected particle luminosity ($\rm erg\ s^{-1}$)\markd	& $L_{\rm inj}$ & $5.4\times 10^{45}$ & $2\times 10^{46}$ & $2\times 10^{46}$  &	$5\times 10^{46}$ \\ \hline 
\end{tabular}}
\end{center}
\vspace{-0.5 mm}
\footnotesize{
{\bf Notes.} Parameters for the SED model for Q0906 in Figure~\ref{fig:fig3}.\\}
$^{\rm a}$ Redshifts from NED. $M_\odot$ and $L_{disk}$ tuned from \citet{gtsg15} to match SED.\\
$^{\rm b}$ $\delta$ range allowed by the SED ($\delta=10$ is assumed in deriving the other parameters). \\
$^{\rm c}$ Effective radius of the elongated jet computed with $R'_E=(3{R'}_b^{2} t_{evol}c/4)^{1/3}$.\\
$^{\rm d}$ Energy injected into the jet in the jet rest frame.\\
\end{table*}

\subsection{Comparison with other $z>5$ Blazars}
\label{sec:sec3_5}

	Since their SEDs are quite similar, we next make a comparative analysis for the
other three claimed $z>5$  blazars B2~1023+25, SDSS~J0131$-$0321, and SDSS~J114657.79+403708.6
\citep[hereafter B2~1023, J0131, and J1146;][]{stgp+13,gstf+14,gtsg15}. These sources
are all radio loud (to varying degrees) and thus represent a population of high-mass,
spin-dominated BHs in the early universe. We can update earlier characterization of
the SEDs around the critical EC peak, by re-measuring the X-ray archival data and by
deriving improved {\it Fermi}-LAT upper limits by using 8 years of Pass-8 data. Our
model also differs from, e.g. \citet{gtsg15} in that we integrate over the cooling population
in the emission zone and that we assign the mm-IR fluxes to the synchrotron peak
(rather than a dust torus component, see Section~\ref{sec:sec4}).

	For these blazars, we used {\it WISE}, {\it Spitzer}, and 2MASS
catalogs for the IR band, and {\it SDSS} catalog and
the GROND data (only for B2~1023) reported in \citet{stgp+13} for the optical band.
Note that these data are not contemporaneous.
We reanalyzed the {\it Chandra}, {\it Swift} and {\it NuSTAR} X-ray data used in the previous studies
whenever available, and further included new 8-year {\it Fermi}-LAT data for the higher energies.
The X-ray and {\it Fermi}-LAT data are processed and analyzed as for Q0906.
We show the broadband SEDs of B2~1023, J0131, and J1146 in Figure~\ref{fig:fig3}.
We model these SEDs using our synchro-Compton model below. Note that we do not have IR spectroscopic
data for these blazars for estimating their masses, so we start from the $M_\bullet$ estimates
of \citet{gtsg15}, adjusting as needed to match the optical/IR SEDs.

\begin{figure}
\centering
\hspace{-7 mm}
\begin{tabular}{cc}
\includegraphics[width=3.5 in]{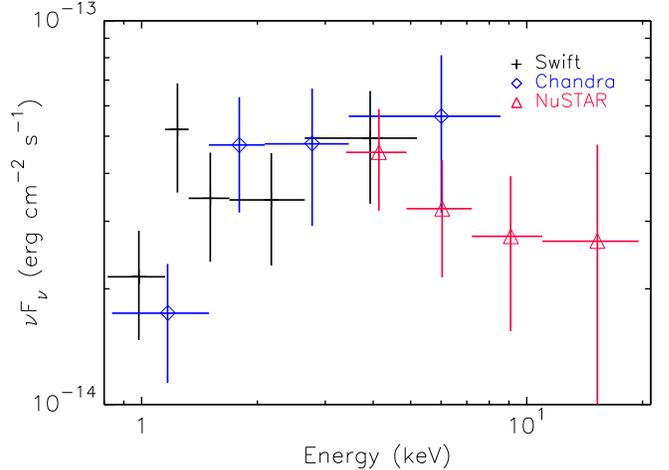} 
\end{tabular}
\figcaption{X-ray SED of B21023 measured with the archival {\it Swift},
{\it Chandra} and {\it NuSTAR} data.
\label{fig:fig5}
}
\vspace{0mm}
\end{figure}
	 For B2~1023, the overall IR-to-X-ray SED we construct is very similar to that reported
previously \citep{stgp+13,gstf+14,gtsg15}. However our LAT upper limits improve
by over $10\times$ and this rules out the highest-$\Gamma_{\rm D}$ or the smallest-$\theta_{\rm V}$ 
models in Figures~2 and 3 of \citet{stgp+13}. Moreover our re-analysis of the
{\it NuSTAR} spectrum (using $R=30''$ and $R=45''$ apertures for source and background
extraction, respectively) does not agree with their finding of a steeply rising flux (Figure~\ref{fig:fig5}).
Instead we see a break similar to that of Q0906 (Figure~\ref{fig:fig2}). This seems a true
discrepancy in the analysis: using their reported spectral parameters {\tt WebPIMMS} gives 
combined expected 4--20\,keV count for HPD extraction from the two {\it NuSTAR} modules of
200 for their $\Gamma=1.29^{+0.14}_{-0.15}$
($F_{\rm 5-10\,keV}=5.8\times10^{-14}\rm \ erg\ s^{-1}\ cm^{-2}$)
model and 190 counts if $\Gamma=1.60^{+0.27}_{-0.26}$
($F_{\rm 5-10\,keV}=5.5\times10^{-14}\rm \ erg\ s^{-1}\ cm^{-2}$).
However \citet[][and we here]{stgp+13}
find only 90 detected counts. Comparing with the results for our simple power-law fit
($\Gamma=1.5\pm0.2$ and $F_{\rm 5-10\rm \ keV}=2.5\pm0.5\times 10^{-14}\rm \ erg\ s^{-1}\ cm^{-2}$
with cross-normalization factors of $\sim$2)
we predict 90 events in good agreement with the observations. Hence a 4--20\,keV extension of 
their rather hard inferred spectrum is difficult to accommodate.
We tested additional changes to the source aperture center ($d=10''$) and size ($R=20''$),
and the location of background extraction region to see if this result is sensitive to the
data selection; all fits values remain consistent with those reported above and
inconsistent with those of \citet{stgp+13}.
Part of this discrepancy might be
the 15\% correction to the {\it NuSTAR} effective area inferred since
CALDB 20131007\footnote{http://heasarc.gsfc.nasa.gov/docs/heasarc/caldb/nustar/docs\\/release\_20131007.txt},
but this does not explain the full $2\times$ discrepancy. Possibly they renormalized
the {\it NuSTAR} flux in a joint fit. If we follow the \citet{stgp+13} binning to one count/bin
and analyze with  the {\tt cstat} statistic, we find a joint fit requires a very large
cross-normalization factors of $\sim$1.8--2. This might be accommodated with a large source variability,
but this is not supported by the {\it CXO} and {\it Swift} data so we consider this improbable.

	Over all, B2~1023 can be fit with parameters rather similar to Q0906,
although we require small modifications to the disk temperature and luminosity and
the synchrotron/SSC normalization (Table~\ref{ta:ta2}). The relatively poor X-ray S/N
at the SCC peak and the lack of mid-IR detection allow a larger $\delta$ range.
The {\it Fermi} bounds still limit $\delta<28$, but on the low side we can
accommodate $\delta$ as small as 4. Measured $\sim10^{12}-10^{14}$\,Hz fluxes 
would help, lowering $\delta_{max}$, while a deeper {\it NuSTAR} exposure can pin down the
typical $>10$\,keV flux level to tighten up $\delta_{min}$. For example, if the
soft X-ray spectrum of \citet{stgp+13} really does continue, we require 
higher electron energies (e.g., $\gamma_{\rm e}\sim1.5\times 10^3$) and can accommodate
$\delta \lapp 3$ with some {\it NuSTAR} flux contributed by EC emission (see Figure~\ref{fig:fig3}).
Note that with more limited SED coverage we did not attempt X-ray $\chi^2$ optimization 
of the model parameters as for Q0906 (Section~\ref{sec:sec3_4}).

	The SEDs of J0131 and J1146 are even less well measured, but do show some
differences from the other two. J0131 has very strong thermal
disk emission and J1146 has relatively small SSC flux compared to its synchrotron emission.
For the these blazars low frequency IR points are useful for estimating the synchrotron 
component flux, but the lack of hard X-ray measurements leaves the SCC peak
frequency almost unconstrained. Thus $\gamma_{\rm e}\approx(\nu_{ssc,pk}^{obs}/\nu_{sy,pk}^{obs})^{1/2}$
is similarly unconstrained. The improved LAT upper limits from our analysis do 
place a bound $\delta \lapp 16$ in both cases, but $\delta$ as small as 4 seems acceptable
(Figure~\ref{fig:fig3} bottom).
The model parameters for the $\delta=10$ case are presented in Table~\ref{ta:ta2}.
Note that for J0131, we need a large magnetic field to prevent the EC emission
from intruding on the {\it Fermi} upper limits for the given synchrotron amplitude.
The model parameters we infer (Table~\ref{ta:ta2}) are similar to those reported in \citet{ppfs16}
for other high-$z$ blazars ($z$=2.4--4.7).

\section{Discussion and Conclusions}
\label{sec:sec4}

	We have analyzed new data for Q0906 taken with Gemini, {\it NuSTAR} and
{\it Fermi}-LAT measurements.  Our check of archival {\it Swift} exposures implies
that the blazar's X-ray emission at our new epoch is quite consistent with historical values.
Indeed Q0906 has been quite constant for over 10 years and so the gamma-ray data can 
be averaged over the 8-yr LAT data set and combined with archival radio, IR and optical data
to assemble a broad-band SED of Q0906.  The Gemini spectra also provide a
$\sim$4$\times10^9 M_\odot$ virial mass estimate for the BH. 

	In our {\it Fermi}-LAT data analysis, we do not detect Q0906, with a 95\% flux upper
limit in the $\sim$GeV band $\sim 10^{-13}\rm \ erg\ s^{-1}$. This is approximately 
two orders of magnitude lower than that implied by the {\it EGRET} excess counts. We thus infer that
the excess was most likely a statistical fluctuation. Alternatively it could represent
decadal-scale variability with a very bright (and soft) flare at the {\it EGRET} epoch.
The {\it NuSTAR} X-ray data indicate a hard X-ray break, so that the peak of the
X-ray emission, identified with the SSC component, is below 10\,keV. This places
a lower limit on $\delta$ so that the SSC peak matches the {\it NuSTAR} data and
the EC up-scattered disk emission does not intrude on the 0.5--79\,keV {\it NuSTAR} band,
while the synchrotron component still explains the {\it Spitzer} flux.
Thus our new measurements bound $6 < \delta < 11.5$ using our synchro-Compton model. 

	Similar and variable hard X-ray breaks have been seen in other
blazars \citep[][]{hnms+15,tgph+15,sgtp+16,ppfs16} and have been interpreted as
an intrinsic curvature of the high-energy emission \citep[EC or SSC;][]{tgph+15,hnms+15}.
Our interpretation of the break is similar to that of \citet{hnms+15};
the break is seen because the SSC peak is in the X-ray band \citep[e.g., see Fig.~9 of][]{hnms+15}.
Variability, although not seen in Q0906,  can be explained
by the variation of EC emission or $\delta$; if EC emission becomes stronger or $\delta$ lowers,
the hard X-ray spectrum may become harder, extrapolating well from the low energy index.

	On the observational side the $\delta$ range can be tightened if deeper {\it NuSTAR} or
{\it XMM-Newton} observations refine measurement of our estimated 4\,keV spectral break
and $\nu_{ssc,pk}^{obs}$. Additional far-IR and sub-mm observations constrain better
$\nu_{sy,pk}^{obs}$.  In particular these can distinguish synchrotron emission
(assumed dominant here) from thermal emission from a dust torus, as assumed for this band
by e.g. \citet{gtsg15}.  Other observations can also help. For example, \citet{zafg+17} 
estimated $\delta \approx 4$ for Q0906 from radio brightness temperatures. This
applies to larger radius where Compton drag should reduce $\delta$, but such measurements
can at least provide an independent lower limit to $\delta$ at the jet base. On the modeling side,
we note that we have assumed a jet base $h=0.03$\,pc from the black hole. If this is
larger the EC flux from up-scattered disk photons can be reduced since the seed photon density
scales as $h^{-2}$ once $h$ exceeds the characteristic disk scale. For Q0906 this makes
little difference for the allowed $\delta$ range, but it can allow larger
$\delta_{max}$ for other sources where the LAT upper limits provide the effective bound.

	With our new range on $\delta$ we can make inferences about the source population.
The substantial $\delta_{min} = 6$ means that the viewing angle $\theta_{\rm V}$ should 
be less than $\theta_{max} = \mathrm{cos}^{-1}(\sqrt{1-\delta_{min}^{-2}}) = 9.6 ^\circ$, and the
chance probability to get a source seen at $\lapp$9.6$^\circ$ is only $\lapp$1.4\%;
$\gapp$70 similar high-mass high-$a$ BHs at a similar redshift are expected.
If the true $\delta$ is larger this number increases.
 If we assume a distribution
$\delta^{-s}$ we can compute that the fraction of all blazars seen is ($s\ne1$)
$$f_B = \left[ \frac{1}{2} -\frac{(1-s)}{2(\delta_M^{1-s} - \delta_m^{1-s})}\int_{\alpha_m}^{\alpha_M} {\rm sin}^2 x {\rm cos}^{s-2} x dx
\right ]\times 2,$$
where $\alpha_M = {\rm cos}^{-1} 1/\delta_M$, $\alpha_m = {\rm cos}^{-1} 1/\delta_m$, and the 
factor 2 at the end assumes a similar jet and counter-jet.
For a uniform prior ($s=0$) this is
$$f_B = 1 - \frac{1}{(\delta_M - \delta_m)}\left (\alpha_m-\alpha_M + {\rm tan}\alpha_M - {\rm tan}\alpha_m\right )
$$
which gives 140 unobserved blazars like Q0906 for $\delta =6-11.5$. With our weaker constraints
for the other sources we obtain 230 blazars like B2 1023 and 130 each like J0131 and J1146.
Of course we do not know if these four objects represent a complete sample of the 
$z>5$ blazar population beamed toward Earth. The targets are drawn from radio surveys so in principle
areal densities of similar objects could be computed. However the completeness of the
follow-up SED and spectral observations that qualify them as blazars is less certain. 
Also luminosity bias associated with Doppler boosting might weight the detection probability
over the allowed $\delta$ range. For example \citet{asrd+12} infer a power-law distribution
with $s=2$. In this case we find
$$ f_B = 1 -  \frac{\delta_M \delta_m}{2(\delta_M - \delta_m)}\left (\alpha_M-\alpha_m -\frac{{\rm sin}2\alpha_M - {\rm sin}2\alpha_m}{2}\right ) .$$
In this case we get 120 (Q0906-like), 80 (B2 1023-like) and 70 (each J0131 and J1146-like).
A detailed treatment of the selection effects goes beyond the present paper, but conservatively
interpreting our sample as complete, we see that the four $z>5$ objects represent
a population of 620 (for uniform prior) or 350 (for $s=2$ prior)
high-mass (hence luminous), high-spin (hence jet-dominated) black holes at this large redshift.

	Such large numbers are interesting since from the optical SDSS survey-derived
black hole mass functions in \citet{vo09} we can estimate a volume density of 
$\sim 15 {\rm Gpc^{-3}}$ $M_\bullet > 10^9 M_\odot$ active black holes in the redshift range
$z=4.3\pm0.7$.
The emission detected by SDSS (optical SED and emission line detections) is nearly isotropic and
the density evolution in \citet{vo09}'s highest $z$ bins appears slow, so from this we estimate 
3150 massive AGN in the 210 Gpc$^3$ between $z=5$ and $z=5.5$.
Thus our radio-loud blazars represent an estimated 10--20\%
of this population. Incompleteness
of the radio blazar IDs would increase the fraction; decreased $\delta$ would lower it.
But the main conclusion, that a very substantial fraction of bright high-$z$ AGN are jet dominated,
seems firm.

	\citet{bv08} have studied spin distributions of BHs using numerical simulations.
They focused on three cases for growth of BHs and found that depending on the growth
process the final spin distribution differs. In particular, only when BHs grow
with prolonged accretion there can be a significant number of high-spin black
holes at $z>5$; in the cases that BHs grow via mergers or chaotic accretion only,
not many BHs are expected to have large spin. Thus our SED-mediated population estimate
suggests that many massive black holes had significant early disk accretion, and have been
driven to high angular momentum $a$. As emphasized by \cite{gtsg15}, such high $a$ means
high total accretion luminosity $\eta$ and, for a given Eddington flux, a lower value
for the total mass accretion rate. In turn that means high BH masses at $z>5$, such 
as the $M_\bullet \approx 1.5 \times 10^{10}M_\odot$ inferred for J0131, are very
hard to achieve at such early times. Perhaps, as suggested by these authors, the very
jet (which is drawing down the black hole spin energy) serves to entrain and
redirect part of the accretion luminosity, allowing a larger accretion rate and
faster black hole growth. Improved SED observations and modeling of these rare, but
demographically important high-$z$ blazars remains the key to probing this early
back hole evolution.
\bigskip

The \textit{Fermi} LAT Collaboration acknowledges generous ongoing support
from a number of agencies and institutes that have supported both the
development and the operation of the LAT as well as scientific data analysis.
These include the National Aeronautics and Space Administration and the
Department of Energy in the United States, the Commissariat \`a l'Energie Atomique
and the Centre National de la Recherche Scientifique / Institut National de Physique
Nucl\'eaire et de Physique des Particules in France, the Agenzia Spaziale Italiana
and the Istituto Nazionale di Fisica Nucleare in Italy, the Ministry of Education,
Culture, Sports, Science and Technology (MEXT), High Energy Accelerator Research
Organization (KEK) and Japan Aerospace Exploration Agency (JAXA) in Japan, and
the K.~A.~Wallenberg Foundation, the Swedish Research Council and the
Swedish National Space Board in Sweden.
 
Additional support for science analysis during the operations phase is gratefully
acknowledged from the Istituto Nazionale di Astrofisica in Italy and the Centre
National d'\'Etudes Spatiales in France. This work performed in part under DOE
Contract DE-AC02-76SF00515.

This work was supported in part by NASA grant NNX17AC27G under
the NuSTAR guest observer program.
This research was supported by Basic Science Research Program through
the National Research Foundation of Korea (NRF)
funded by the Ministry of Science, ICT \& Future Planning (NRF-2017R1C1B2004566).

\bibliographystyle{apj}
\bibliography{GBINARY,BLLacs,PSRBINARY,PWN,STATISTICS,FERMIBASE,ABSORB}

\end{document}